\documentclass[twocolumn,notitlepage,showkeys,showpacs,prl]{revtex4-1}
\usepackage{graphicx}
\usepackage{dcolumn}
\usepackage{booktabs,bm,color,braket,amsmath}
\usepackage{txfonts}
\usepackage{ulem}
\usepackage{color}
\usepackage[dvipsnames]{xcolor}
\usepackage[colorlinks,linkcolor=blue,citecolor=blue,urlcolor=blue]{hyperref}

\def\add#1{\textcolor{blue}{#1}}

\def\bs#1{\boldsymbol{#1}}

\begin{document}

\title{Higher-order, quantum \add{magnetic} inductions in chiral topological materials}

\author{Yizhou Liu}
\email{yizhou.liu@weizmann.ac.il}
\author{Hengxin Tan}
\author{Binghai Yan}
\affiliation{Department of Condensed Matter Physics, Weizmann Institute of Science, Rehovot 7610001, Israel}

\date{\today}

\begin{abstract}
The classic magnetic induction effect is usually considered in electric circuits or conductor coils. In this work, we propose quantum induction effects induced by the Berry curvature in homogenous solids. Two different types of quantum inductions are identified in the higher order of a magnetic field. They are closely related to the magnetic-field-induced anomalous velocity proportional to the Berry curvature so that both quantum inductions become significantly enhanced near Weyl points. Further, we propose an ac-magnetic-field measurement to detect quantum induction effects and distinguish them from classical induction.
\end{abstract}

\maketitle

\textit{Introduction}.---Faraday's law of induction, which describes the current generated by a varying magnetic field, is usually considered in the coil geometry. But it also exists in uniform gyrotropic materials as the low frequency limit of optical gyrotropy\cite{landau2013electrodynamics} and gains renewed interest in recent studies of topological materials.

In Weyl semimetals\cite{Armitage2018RMP,Yan2017}, the chiral magnetic effect (CME) known previously in particle physics\cite{Fukushima2008CME} can lead to a nonequilibrium current in the presence of a nonorthogonal electric field and magnetic field ($\bs{B}$)\cite{Son2012CME}. The CME roots in the chiral anomaly and the Berry curvature of Weyl fermions\cite{Nielsen1983,Son2013PRB}. \add{A similar effect} was also proposed to generate current by a varying magnetic field ($\bs{\dot{B}}$) rather than the static $\bs{B}$ \cite{Zyuzin2012PRB,Vazifeh2013PRL, Zhou2013CPL, Bassar2014PRB,Goswami2015}. This dynamical \add{effect} was later recognized as a gyrotropic magnetic effect (GME)\cite{Zhong2016PRL}. It is equivalently a Faraday induction effect and not governed by the chiral anomaly or Berry curvature.

In this work, we find that the Faraday induction includes quantum contributions in the higher order in addition to GME. The quantum induction (QI) effects are intimately induced by the Berry curvature in general and related to the chiral anomaly for a Weyl semimetal. They can be significantly enhanced near Berry curvature monopoles such as Weyl points or multiply-degenerate points. We have calculated QIs in chiral topological semimetals CoSi and MnSi and proposed experiments to distinguish QIs from the coexisting classical induction (CI).

\textit{Semiclassical theory of induction}.---According to the semiclassical equation of motion of electrons \cite{Chang1996PRB, Sundaram1999PRB, Xiao2010RMP} the electronic velocity $\dot{\bs{r}}$ under external magnetic field $\bs{B}(t)$ can be derived as,
\begin{gather}
D_{\bs{k}} \dot{\bs{r}} = \tilde{\bs{v}}_{\bs{k}} + \frac{e}{\hbar} (\tilde{\bs{v}}_{\bs{k}} \cdot \bs{\Omega_k}) \bs{B}, \label{velocity} \\
\tilde{\bs{v}}_{\bs{k}} = \bs{v_k} - \frac{1}{\hbar} \nabla_{\bs{k}} (\bs{m_k} \cdot \bs{B}), \label{vk}
\end{gather}
with $D_{\bs{k}} = 1 + \frac{e}{\hbar}\bs{\Omega_k} \cdot \bs{B}$ being the modified density of states \cite{Xiao2005PRL}, $\bs{v}_{\bs{k}}$ refers to the ordinary band velocity $\bs{v_k}=\frac{1}{\hbar} \nabla_{\bs{k}} \varepsilon_{\bs{k}}$, and $\bs{\Omega_k}$ and $\bs{m_k}$ are the Berry curvature and orbital magnetic moment, respectively. The second term on the right-hand side of Eq. \eqref{velocity} is the magnetic-field-induced anomalous velocity which gives rise to the chiral anomaly in the presence of an extra electric field \cite{Son2013PRB}.

In this work, we focus on the effect of magnetic-field-induced current. The electric current is expressed as
\begin{equation}\label{j_tot}
\bs{j} = -e\int \frac{\text{d}\bs{k}}{(2\pi)^3} D_{\bs{k}} \dot{\bs{r}} f,
\end{equation}
with $f$ being the nonequilibrium distribution function of electrons which the Boltzmann transport equation can determine in the relaxation time approximation:
\begin{equation}
- \frac{f - f_0}{\tau} = \dot{\bs{k}} \cdot \nabla_{\bs{k}} f + \dot{\bs{r}} \cdot \nabla_{\bs{r}} f + \partial_t f,
\end{equation}
where $f_0 = \frac{1}{e^{(\varepsilon - \varepsilon_F)/k_BT}+1}$ is the equilibrium Fermi-Dirac distribution function with $\varepsilon_F$ being the Fermi energy. For uniform but time-dependent magnetic field $\bs{B}(t)$, the nonequilibrium distribution function $f$ up to the first order of relaxation time $\tau$ can be derived as
\begin{equation}\label{f}
f = f_0 + \tau (\bs{m_k} \cdot \dot{\bs{B}}) \left.\frac{\partial f_0}{\partial \varepsilon}\right|_{\varepsilon=\varepsilon_{\bs{k}}} + O(\tau^2),
\end{equation}
where $\bs{m_k}\equiv -\frac{\partial \varepsilon}{\partial \bs{B}}$. By substituting Eqs. \eqref{velocity}, \eqref{vk}, and \eqref{f} into \eqref{j_tot}, we get the final expression of the induction current as
\begin{align}
\bs{j} =& \bs{j}^c + \bs{j}^q, \label{j_final} \\
\bs{j}^c =& -e\tau \int \frac{\text{d}\bs{k}}{(2\pi)^3} (\bs{m_k} \cdot \dot{\bs{B}}) \bs{v_k} \left. \frac{\partial f_0}{\partial \varepsilon} \right|_{\varepsilon=\varepsilon_{\bs{k}}} = \sum_{\alpha,\beta=x,y,z} C_{\alpha\beta} \dot{B}_\beta \bs{e}_\alpha,\label{jc_final} \\
\bs{j}^q =& -\frac{e^2\tau}{\hbar} \bs{B} \int \frac{\text{d}\bs{k}}{(2\pi)^3} (\tilde{\bs{v}}_{\bs{k}} \cdot \bs{\Omega_k}) (\bs{m_k} \cdot \dot{\bs{B}}) \left. \frac{\partial f_0}{\partial \varepsilon} \right|_{\varepsilon=\varepsilon_{\bs{k}}} \\
=& \left( \bs{Q}^{(1)} \cdot \bs{\dot{B}} \right) \bs{B} + \left( \sum_{\alpha,\beta=x,y,z} Q^{(2)}_{\alpha\beta} \dot{B}_\alpha B_\beta \right) \bs{B}. \label{jq_final}
\end{align}
Equations \eqref{j_final}--\eqref{jq_final} are the central results of this work. The total induction current $\bs{j}$ can be divided into two parts, $\bs{j}^c$ and $\bs{j}^q$, to represent the classical induction and quantum induction, respectively. Here, $\bs{j}^q$ explicitly contains the Planck constant $\hbar$ and also the Berry curvature $\bs{\Omega}_{\bs{k}}$. The classical term can be rewritten into a compact form of $j^c_\alpha = \sum_{\beta} C_{\alpha\beta} \dot{B}_\beta$ ($\alpha, \beta = x, y, z$) describing conventional Faraday electromagnetic induction in materials. The coefficients are
\begin{equation}\label{C}
C_{\alpha\beta} = -e\tau \int \frac{\text{d}\bs{k}}{(2\pi)^3} v_\alpha m_\beta \left. \frac{\partial f_0}{\partial \varepsilon} \right|_{\varepsilon=\varepsilon_{\bs{k}}},
\end{equation}
which was consistent with previous formulations as the low-frequency limit of natural optical activity, or natural gyrotropy\cite{Goswami2015,Zhong2016PRL}. Corresponding inverse effect is the current-induced magnetization is described by $\bs{M} = C^T \bs{E}$ ($T$ stands for matrix transpose) \cite{Zhong2016PRL, Yoda2018NanoLett} with $\bs{E}$ and $\bs{M}$ being the electric field and induced magnetization, respectively. According to Eq.~\eqref{C}, the diagonal coefficients ($C_{\alpha\alpha}$) vanish in the presence of inversion symmetry or mirror symmetry. Therefore, the ordinary induction can only occur in chiral systems \cite{Yoda2015SciRep} or solenoids \cite{Yoda2018NanoLett}.

The quantum induction current $\bs{j}^q$ is higher order in $\bs{B}$ and $\bs{\dot{B}}$. Up to our knowledge, it is not discussed in literature so far. It can be further divided into two parts as: $\bs{j}^q = \left( \bs{Q}^{(1)} \cdot \dot{\bs{B}} \right) \bs{B} + \left( \sum_{\alpha\beta} Q^{(2)}_{\alpha\beta} \dot{B}_\alpha B_\beta \right) \bs{B}$ with $\bs{Q}^{(1)}=(Q^{(1)}_x, Q^{(1)}_y, Q^{(1)}_z)$ and $Q^{(2)}_{\alpha\beta}$ being the response coefficients of QIs whose expressions are given by:
\begin{equation}
\begin{split}\label{L1_and_L2}
Q^{(1)}_\alpha =& - \frac{e^2\tau}{\hbar} \int \frac{\text{d}\bs{k}}{(2\pi)^3} (\bs{v_k} \cdot \bs{\Omega_k}) m_\alpha \left.\frac{\partial f_0}{\partial \varepsilon}\right|_{\varepsilon=\varepsilon_{\bs{k}}} \\
Q^{(2)}_{\alpha\beta} =& - \frac{e^2\tau}{\hbar^2} \int \frac{\text{d}\bs{k}}{(2\pi)^3} m_\alpha \left( \bs{\Omega_k} \cdot \nabla_{\bs{k}} \right) m_\beta \left. \frac{\partial f_0}{\partial \varepsilon} \right|_{\varepsilon=\varepsilon_{\bs{k}}}.
\end{split}
\end{equation}
Recall $\bs{v_k}$ and $\nabla_{\bs{k}}$ are odd under inversion symmetry ($\mathcal{P}$) or time-reversal symmetry ($\mathcal{T}$) while $\bs{\Omega_k}$ and $\bs{m_k}$ are even under $\mathcal{P}$ but odd under $\mathcal{T}$. $Q^{(1)}_\alpha$ exists when both $\mathcal{P}$ and $\mathcal{T}$ are broken. $Q^{(2)}_{\alpha\beta}$ only requires $\mathcal{P}$-breaking, similar to $C_{\alpha\beta}$. Because $\bs{j}^q$ is proportional to the Berry curvature $\bs{\Omega_k}$, it is naturally expected that the quantum induction is enhanced in $\mathcal{P}$-breaking topological semimetals such as the Weyl semimetal or triple-point semimetal. The Berry curvature contribution in Eq.\eqref{C} has the same origin as the chiral anomaly effect.

\textit{Weyl semimetal}.---Take the Weyl semimetal as an example. We can estimate the magnitudes of the coefficients $C$, $Q^{(1)}$, and $Q^{(2)}$. Near the Weyl point, the group velocity $v$ is a constant while the Berry curvature and the orbital magnetic moment scale as $\Omega \propto \frac{1}{k^2_F}$ and $m \propto \frac{1}{k_F}$, respectively, where $k_F = \frac{\varepsilon_F - \varepsilon_W}{\hbar v_F}$ is the Fermi wave vector with $\varepsilon_W$ and $v_F$ being the energy of Weyl point and Fermi velocity, respectively. Therefore, we obtain $C \propto k_F$ for a single Weyl point and $C \propto (\varepsilon_F - \varepsilon_{W1}) - (\varepsilon_F - \varepsilon_{W2})=\varepsilon_{W2}-\varepsilon_{W1}$ for a pair of Weyl points. Here $C$ depends on the energy separation of Weyl points but less sensitive to $\varepsilon_F$. In contrast, $Q^{(1)} \propto \frac{1}{k_F}$ and $Q^{(2)} \propto \frac{1}{k^3_F}$ become divergent when the Fermi energy approaches Weyl points.

\begin{figure}
\centering
\includegraphics[width=\linewidth]{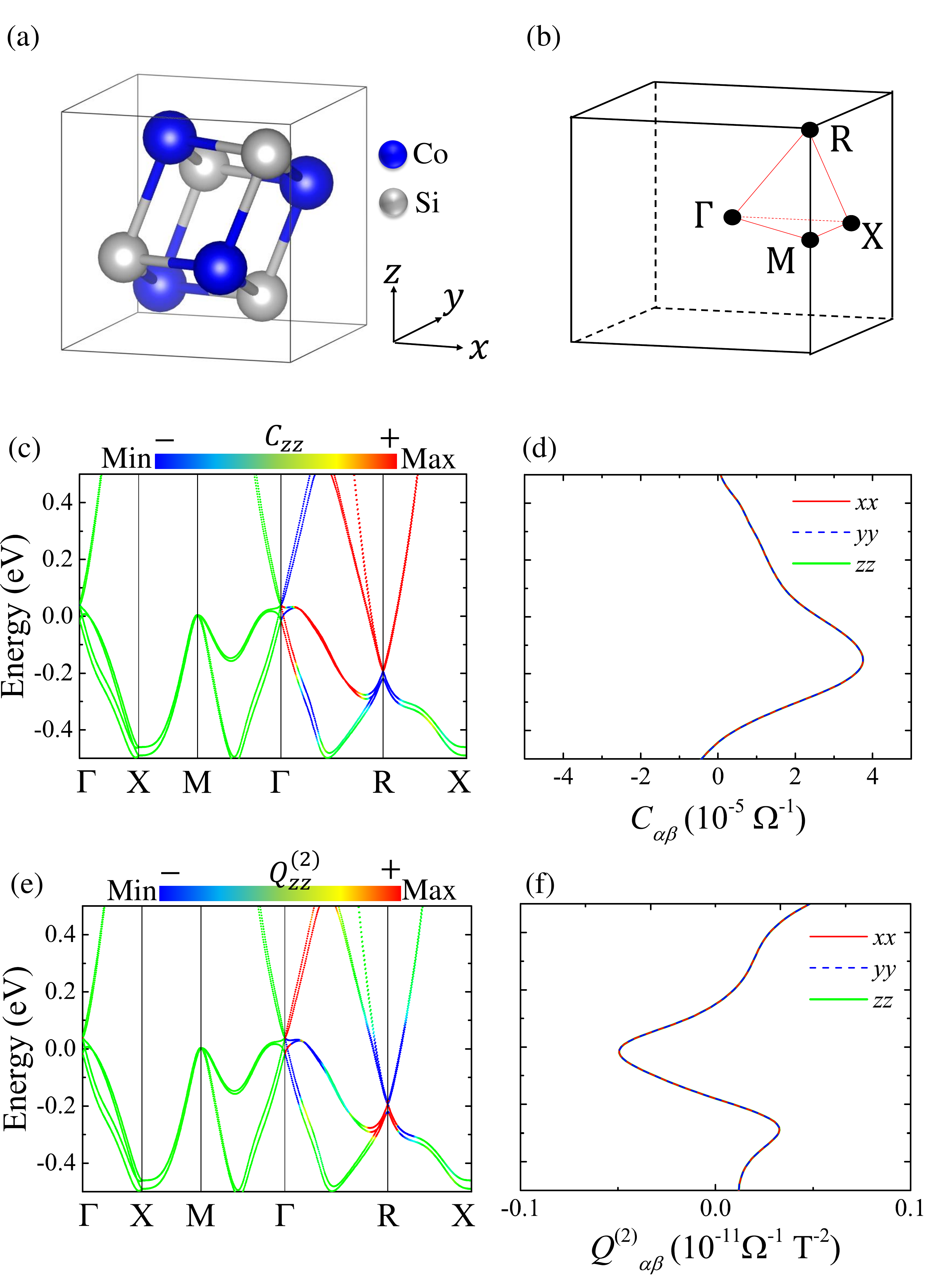}
\caption{\label{fig1}
\textbf{CI and QI in CoSi.} (a) Chiral crystal structure of CoSi with cubic point group symmetry $T$. (b) Brillouin zone (BZ) and high-symmetry lines. (c) Bulk band structure and CI of CoSi. (d) $C_{\alpha\alpha}$ as function of Fermi energy $\varepsilon_F$. (e) Bulk band structure and second QI of CoSi. (f) $Q^{(2)}_{\alpha\alpha}$ as function of $\varepsilon_F$. The first QI $Q^{(1)}_\alpha=0$ due to TRS.
}
\end{figure}

We consider a minimal model of a Weyl semimetal without $\mathcal{P}$ or $\mathcal{T}$. A minimal model of Weyl Hamiltonian without $\mathcal{T}$ has \textit{one} pair of Weyl points. The inversion (and any mirror) symmetry is naturally broken if the energies of two Weyl points are different, which motivates us to consider the following two-band model:
\begin{equation}
\begin{split}\label{Toy_model}
&H_{\bs{k}} = \varepsilon_0 + d_x \sigma_x + d_y \sigma_y + d_z \sigma_z, \\
&\varepsilon_0 = A_0 \sin k_z, ~~~ d_x = A \sin k_x, ~~~ d_y = A \sin k_y, \\
&d_z = M - B_1 (\cos k_x + \cos k_y) - B_2 \cos k_z,
\end{split}
\end{equation}
with $A_0$, $A$, $B_1$, $B_2$, and $M$ being constants. $\sigma_{x,y,z}$ refer to Pauli matrix. The band structure of such model is given by $\varepsilon_{s\bs{k}} = \varepsilon_0 + s\sqrt{d^2_x + d^2_y + d^2_z}$ ($s=\pm1$ refers to the upper or lower band) with the band crossing happens when $d_x = d_y = d_z =0$. Within the parameters range $|M - 2B_1|<|B_2|<|M|$, $|B_2|<|M+2B_1|$, the band structure has a pair of Weyl points with opposite chiralities on the $k_z$-axis which are determined by $M-B_2\cos k_z =0$. The energies of the two Weyl points become different when $A_0 \ne 0$, and the low energy effective Hamiltonian near the Weyl points become:
\begin{equation}
H_\chi = \hbar (v_1 k_x \sigma_x + v_1 k_y \sigma_y + \chi v_2 k_z \sigma_z) + \varepsilon_\chi + \chi \hbar v' k_z. \label{h_pm}
\end{equation}
with $v_1$ being the group velocity along $k_x$ and $k_y$ axis and $v_2$ the group velocity along $k_z$; $v'$ represents a tilting parameter of the Weyl cone along $k_z$ axis; $\chi=\pm$ refers the Weyl cone with positive (negative) chirality; $\varepsilon_+$ and $\varepsilon_-$ are the energies of Weyl points. Based on the approximation $\left.\frac{\partial f_0}{\partial \varepsilon}\right|_{\varepsilon=\varepsilon_{\bs{k}}} \approx - \delta(\varepsilon_{\bs{k}} - \varepsilon_F)$, the classical coefficients are calculated (for small tilted parameter $v'$) as \cite{SM}:
\begin{equation}\label{C_Effective_model}
C_{\alpha\alpha} \approx \frac{e^2\tau \Delta \varepsilon}{12\pi^2\hbar^2}, ~~~ \alpha=x, y, z
\end{equation}
which depends only on the relaxation time $\tau$ and the energy difference between two Weyl points $\Delta\varepsilon = \varepsilon_{+} - \varepsilon_{-}$. Since $\Delta\varepsilon$ changes sign under inversion or mirror operations, the CI can only exist in chiral systems, consistent with our previous discussion.

Unlike CI $C_{\alpha\beta}$ which depends only on the energy difference between two Weyl points but independent on the Fermi energy $\varepsilon_F$, the QIs $Q^{(1)}_\alpha$ and $Q^{(2)}_{\alpha\beta}$ significantly depends on $\varepsilon_F$ according to our previous discussion. The QIs of the effective model is calculated as \cite{SM}:
\begin{align}
& Q^{(1)}_z = -\frac{e^3\tau v^2_1 v'}{24\pi^2\hbar v_2} \left( \frac{1}{\varepsilon_F - \varepsilon_+} - \frac{1}{\varepsilon_F - \varepsilon_-} \right), ~~~ Q^{(1)}_{x,y} =0,\label{Q1_Effective_model} \\
& Q^{(2)}_{xx}=Q^{(2)}_{yy} \approx \frac{e^4 v^2_1 v^2_2 \tau}{48\pi^2} \left[ \frac{1}{(\varepsilon_F - \varepsilon_+)^3} - \frac{1}{(\varepsilon_F - \varepsilon_-)^3} \right], \label{Q2xxyy_Effective_model} \\
& Q^{(2)}_{zz} \approx \frac{e^4 v^4_1 \tau}{48\pi^2} \left[ \frac{1}{(\varepsilon_F - \varepsilon_+)^3} - \frac{1}{(\varepsilon_F - \varepsilon_-)^3} \right]. \label{Q2zz_Effective_model}
\end{align}
The quantum coefficients $Q^{(1)}$ and $Q^{(2)}$ become zero when $\varepsilon_+ = \varepsilon_-$, \textit{i.e.} when the inversion symmetry $\mathcal{P}$ is restored. Remarkably, according to Eqs. \eqref{Q1_Effective_model}--\eqref{Q2zz_Effective_model}, the quantum coefficients get maximized when $\varepsilon_F$ approaches one of the Weyl points. Numerical results of the full lattice model in Eqs. \eqref{Toy_model} is shown in Ref.\onlinecite{SM} which is consistent with the analytical results.

\begin{figure}
\centering
\includegraphics[width=\linewidth]{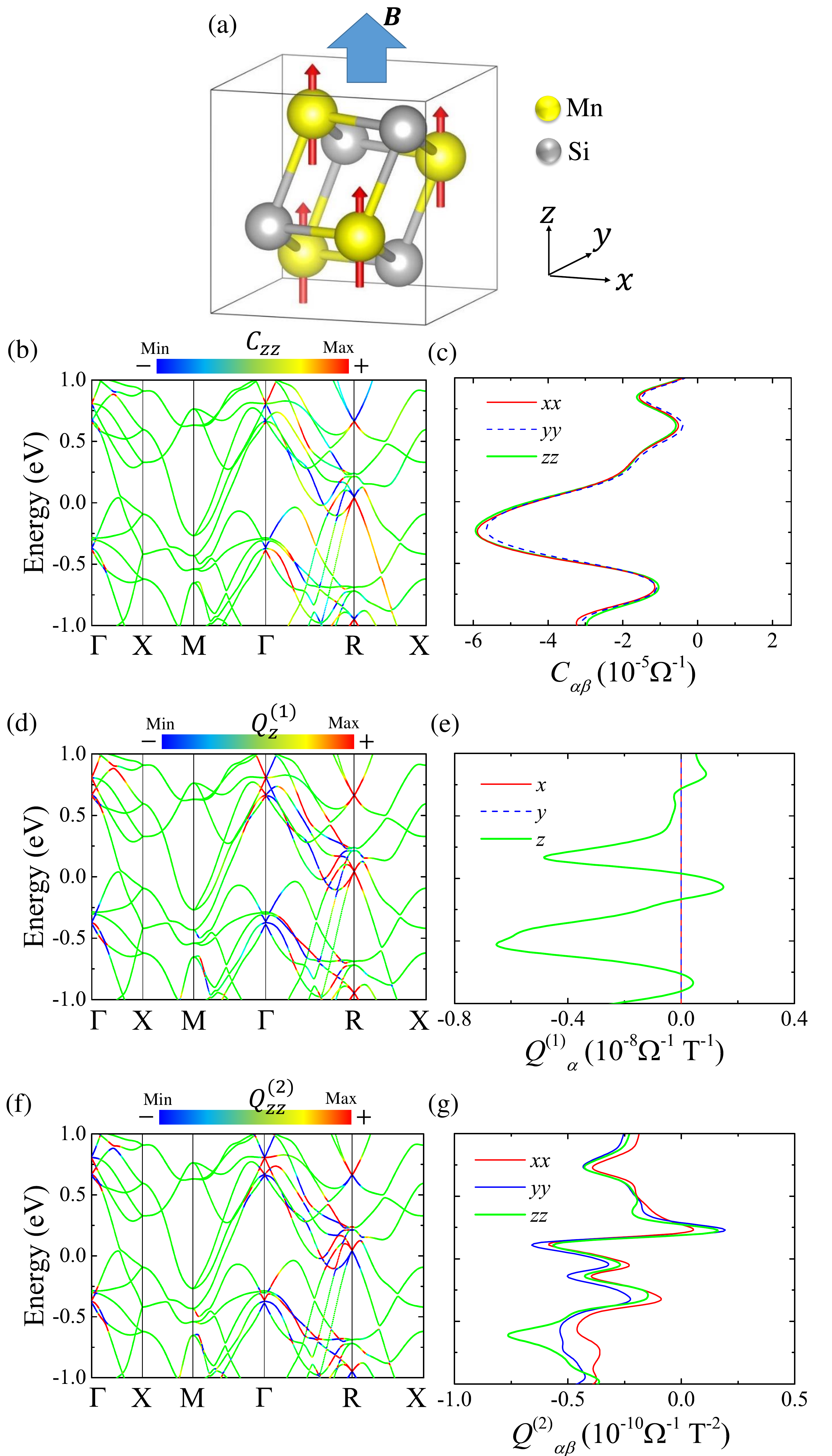}
\caption{\label{fig2}
\textbf{CI and QIs in MnSi.} (a) Crystal and ferromagnetic structure of MnSi under external magnetic field $\bs{B}$. (b) Bulk band structure and CI of ferromagnetic MnSi. (c) $C_{\alpha\alpha}$ as function of $\varepsilon_F$. (d), (f) Bulk band structure and the first (d) and second (f) QI of MnSi. (e), (g) $Q^{(1)}_\alpha$ and $Q^{(2)}_{\alpha\alpha}$ as functions of $\varepsilon_F$. $Q^{(1)}_{x,y}=0$ due to the two-fold (skew) rotation symmetry along $z$-axis.
}
\end{figure}

\textit{Material candidates}.---In the following, we demonstrate the quantum induction effect in inversion-breaking topological materials, among which the chiral topological materials has attracted much research of interest because of the multifold fermion \cite{Chang2017PRL, Chang2018NatMater, Rao2019Nature, Sanchez2019Nature, Takane2019PRL} and the quantized photogalvanic responses \cite{de2017NatCommun}. The transition-metal monosilicides family $X$Si ($X=\text{Rh}, \text{Co}$, etc.) with a chiral cubic crystal structure (Figs. \ref{fig1}(a)-(b)) exhibit a pair of chiral Weyl fermions with large Chern numbers. They have been mostly investigated because the Weyl points' energy is very close to the Fermi level and has long Fermi arcs observed by the angle-resolved photoemission spectroscopy (ARPES). The band structure of CoSi is shown in Fig. \ref{fig1}(c) (see calculation details in Ref. \onlinecite{SM}) with multiple Weyl cones located at $\Gamma$ and R near the Fermi level \cite{Chang2017PRL}.

Because of the cubic symmetry all the off-diagonal components of $C_{\alpha\beta}$ and $Q^{(2)}_{\alpha\beta}$ vanish, and the diagonal coefficients are equal, \textit{i.e.} $C_{xx}=C_{yy}=C_{zz}$, $Q^{(2)}_{xx}=Q^{(2)}_{yy}=Q^{(2)}_{zz}$ and $C_{\alpha\beta}=Q^{(2)}_{\alpha\beta}=0$ ($\alpha\ne\beta$). Figures \ref{fig1}(c)-(d) and \ref{fig1}(e)-(f) show the numerically calculated $C_{\alpha\alpha}$ and $Q^{(2)}_{\alpha\alpha}$, respectively, for a relaxation time $\tau \approx 6.6\times10^{-15}$s (see details of the calculation methods in Ref. \onlinecite{SM}). As is shown $Q^{(2)}_{\alpha\alpha}$ get significantly enhanced at the multiple Weyl points, which is different from $C_{\alpha\alpha}$ of the two-band Weyl Hamiltonian.

In the presence of TRS or cubic symmetry, quantum response $Q^{(1)}$ vanishes. However, both TRS and cubic symmetry can be broken by the magnetization induced by the external magnetic field (Fig. \ref{fig2}(a)). Previous experimental studies have reported ferromagnetic order in the ground state of MnSi at extremely low temperatures \cite{Williams1966JAP, Shinoda1966JPSJ, Wernick1972MaterResBull, Ishikawa1977PRB}.  In the ferromagnetic state, both the TRS and cubic symmetries are naturally broken so that the $Q^{(1)}_\alpha$ can be nonzero. Figures \ref{fig2}(b)-(g) show the band structure and coefficients $C_{\alpha\beta}$, $Q^{(1)}_\alpha$, and $Q^{(2)}_{\alpha\beta}$ of ferromagnetic MnSi (see calculation details in SM \cite{SM}). It should be noted that the $x$- and $y$-components of $Q^{(1)}_\alpha$ vanish because of the two-fold (skew) rotation symmetry along the $z$-axis.

\begin{figure}
\centering
\includegraphics[width=\linewidth]{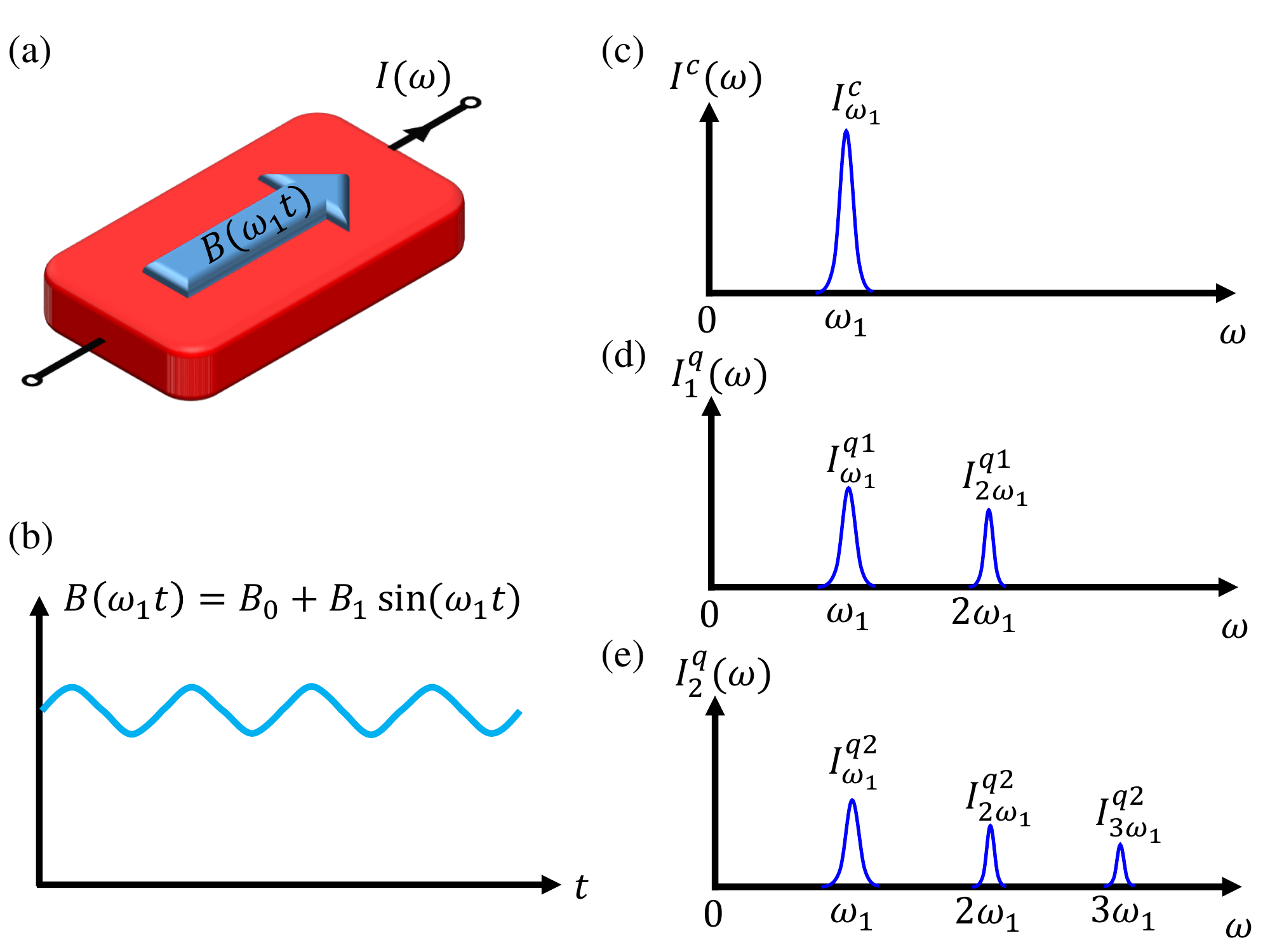}
\caption{
\textbf{Different signatures of CI and QIs in an external ac magnetic field.} (a)-(b) Schematic setup of a sample (red) under an external ac magnetic field $B(\omega_1t)=B_0 + B_1 \sin(\omega_1t)$ having a frequency dependent output electrical current $I(\omega)$. (c)-(e) Schematic of the output signals of classical and quantum induction current $I^c(\omega)$, $I^q_1(\omega)$, and $I^q_2(\omega)$ as function of frequency, respectively. The estimated typical orders of each peak is given in Tab. \ref{tab1}.
}
\label{fig3}
\end{figure}

\textit{Discussion}.---The coexisting classical and quantum induction current can be differentiated in an external ac magnetic field. We consider an ac magnetic field $B(t)=B_0 + B_1 \sin(\omega_1t)$ as schematically shown in Figs. \ref{fig3}(a)-(b). The constant field $B_0$ is used to fix the magnetization of the sample and the time-dependent $B_1\sin(\omega_1t)$ to generate the induction effect. The currents of CI and QIs over a sample with cross-section area $A$, \textit{i.e.} $I^c=C\dot{B}$, $I^q_1=Q^{(1)}\dot{B}B$, and $I^q_2=Q^{(2)}\dot{B}B^2$, is given by:
\begin{equation}\label{Ic_Iq1_Iq2}
\begin{split}
I^c =& I^c_{\omega_1} \cos(\omega_1t), \\
I^q_1 =& I^{q1}_{\omega_1} \cos(\omega_1t) + I^{q1}_{2\omega_1} \sin(2\omega_1t), \\
I^q_2 =& I^{q2}_{\omega_1} \cos(\omega_1t) + I^{q2}_{2\omega_1} \sin(2\omega_1 t) + I^{q2}_{3\omega_1} \cos(3\omega_1t),
\end{split}
\end{equation}
with the expressions of $I^c_{\omega_1}$, $I^{q1}_{\omega_1/2\omega_1}$, and $I^{q2}_{\omega_1/2\omega_1/3\omega_1}$ shown in Tab. \ref{tab1}. Figures \ref{fig3}(c)-(e) show the schematic output of the signals in the frequency domain, but the signals of CI and QIs show different numbers of peaks. We can roughly estimate the orders of magnitude of these peaks in a $\mathcal{T}$-broken chiral Weyl semimetal from Eqs. \eqref{C_Effective_model}, \eqref{Q1_Effective_model}--\eqref{Q2zz_Effective_model}. The scales of energies and velocities are of the order $\sim0.1$eV and $\sim10^{5}$ m/s, respectively, so that for a typical relaxation time $\tau \sim 10^{-14}$ s, and external magnetic field with experimentally achievable parameters $B_0\sim1$ T, $B_1\sim0.1$ T, and $\omega_1\sim100$ kHz \cite{Ma2004JMMM, Jordan1999JMMM}, the current signals are of several orders of pA for $A\sim1$ cm$^2$ as is shown in the last column of Tab. \ref{tab1}. The classical signal dominates at $\omega_1$ but is absent at higher frequencies. Remarkably, the strength of the quantum signal $I^{q1}_{2\omega_1}$ is comparable with the circular photogalvanic effect (CPGE) at low frequency, which is of the order 10--100 pA \cite{de2017NatCommun}. Therefore we believe the QIs can be experimentally observed with currently available techniques.

\begin{table}
\centering
\caption{\label{tab1} Expresions and typical orders of magnitudes of $I^c_{\omega_1}$, $I^{q1}_{\omega_1/2\omega_1}$, and $I^{q2}_{\omega_1/2\omega_1/3\omega_1}$ in Eq. \eqref{Ic_Iq1_Iq2} and Figs. \ref{fig3}(c)-(e) under an external ac magnetic field $B(t)=B_0 + B_1\sin\omega_1t$. Parameters $B_0=1$ T, $B_1=0.1$ T, $\omega_1=100$ kHz are used \cite{Ma2004JMMM, Jordan1999JMMM}. The cross-sectional area is taken as $A=1$ cm$^2$. }
\begin{tabular}{p{1cm}|p{3.5cm}p{3.5cm}}\hline\hline
~                    & Expression                & Typical order (pA) \\\hline
$I^c_{\omega_1}$     & $\omega_1CB_1A$           & $10^6 \sim 10^7$ \\
$I^{q1}_{\omega_1}$  & $\omega_1Q^{(1)}B_0B_1A$  & 10$^3$ \\
$I^{q1}_{2\omega_1}$ & $\frac{\omega_1}{2}Q^{(1)}B^2_1A$   & 10$\sim$100 \\
$I^{q2}_{\omega_1}$  & $\omega_1Q^{(2)}(B^2_0+B^2_1/4)B_1$ & 0.1$\sim$1 \\
$I^{q2}_{2\omega_1}$ & $\omega_1Q^{(2)}B_0B^2_1A$          & 0.01$\sim$0.1 \\
$I^{q2}_{3\omega_1}$ & $-\frac{\omega_1}{4}Q^{(2)}B^3_1A$  & 10$^{-3}\sim10^{-4}$ \\\hline
\end{tabular}
\end{table}

Reference \onlinecite{Yoda2018NanoLett} introduced a dimensionless $\xi$ parameter to describe how ``chiral'' the crystal is based on the inverse GME. It is defined as $\xi=\frac{C_{zz}}{\sigma_{zz}} \frac{c}{S_{xy}}$ where $\sigma_{zz}$ is the electrical conductivity, and $c$ and $S_{xy}$ refer to the length and cross-sectional area of the unit cell along $z$-axis. $\xi$ has the physical meaning of the effective number of helical ``turns'' within a unit cell. Since both $C_{zz}$ and $\sigma_{zz}$ are proportional to relaxation time $\tau$, $\xi$ is an intrinsic material parameter that is independent of $\tau$. For CoSi, the parameter $\xi$ is about 0.3 at the Fermi level and can reach 0.8 by fine-tuning $\varepsilon_F$ \cite{SM}. Reference \onlinecite{Nagaosa2019JJAP} predicted a principle for giant inductance based on the emergent electromagnetic field (EEMF) in a spiral magnet and is recently confirmed by an experimental study \cite{Yokouchi2020Nature}. Our theory is based on the general band structure and does not necessarily require a spiral magnetic structure. Interestingly, MnSi supports helical magnetic structure at low-temperature and low- (magnetic-) field regimes, but ferromagnetic order at higher field strength \cite{Ishikawa1977PRB, Ishikawa1984JPSJ} which makes the material an ideal platform to study different QI effects.

In summary, we proposed two kinds of quantum induction effects in noncentrosymmetric topological materials, proportional to the Berry curvature on the Fermi surface. We illustrate the principle of our theory by studying two chiral topological materials CoSi and MnSi. Furthermore, we propose to observe the quantum induction effect in an ac magnetic field to differentiate it from the classical induction effect. Our findings provide another topological quantum effect induced by the Berry phase in solid-state materials.

\begin{acknowledgements}
We thank Tobias Holder and Daniel Kaplan for their helpful discussions. B.Y. acknowledges the financial support by the Willner Family Leadership Institute for the Weizmann Institute of Science, the Benoziyo Endowment Fund for the Advancement of Science, Ruth and Herman Albert Scholars Program for New Scientists, and the European Research Council (ERC) under the European Union’s Horizon 2020 research and innovation program (Grant No. 815869).
\end{acknowledgements}


%

\end{document}